\documentstyle[11pt]{article}
\font\sqi=cmssq8
\begin{document}
\renewcommand{\thesection}{\Roman{section}}
\rightline{hep-th/9409160}
\rightline{Dubna, JINR-E2-94-342}
\rightline{Vienna, ESI-134 (1994)}
\centerline{\large {SELF-DUAL GRAVITY REVISITED}}
\centerline{Ch. Devchand and V. Ogievetsky}
\centerline{Joint Institute for Nuclear Research, 141980 Dubna, Russia}
\vskip 1 cm
\noindent Abstract:
Reconsidering the harmonic space description of the self-dual Einstein
equations, we streamline the proof that all self-dual pure gravitational
fields allow a local description in terms of an unconstrained analytic
prepotential in harmonic space.
Our formulation yields a simple recipe for constructing self-dual metrics
starting from any explicit choice of such prepotential; and we illustrate
the procedure by producing a metric related to the Taub-NUT solution from
the simplest monomial choice of prepotential.
\section{Introduction}

Many years ago Penrose \cite{penrose} pointed out that the twistor transform
in  flat space remarkably yielded itself to a deformation
to curved space, providing a construction (in principle) of the general
self-dual solution of the Einstein equations.
Several classes of solutions have been explicitly constructed
using the twistor technique (e.g.\cite{ward}).
Further classes of explicit self-dual  metrics have been found
by finding classes of particular solutions to the second-order partial
differential equation to which Plebanski \cite{pleb} reduced the self-dual
Einstein equations (see e.g. the review \cite{bfp} for details of this
approach).

The aim of this paper is to describe a somewhat more tractable version of the
`curved twistor construction' of local solutions using the harmonic space
description. Harmonic spaces were originally devised \cite{har}
as tools for the construction of unconstrained off-shell N = 2 and 3
supersymmetric theories. This involved the `harmonisation' of the internal
unitary automorphism groups $G$ of the Poincar\'e supersymmetry algebra,
i.e. the inclusion of harmonics on some coset of G as auxiliary variables;
quantities in conventional superspace being recoverable as coefficients
in a harmonic decomposition. Subsequent applications have involved
harmonic spaces in which the rotation group (rather than some internal
symmetry group) is `harmonised', i.e. these
harmonic spaces are cosets of the  Poincar\'e group by a  subgroup $H$
of the rotation group. Ordinary four-dimensional space, recall,
is the coset of the Poincar\'e group by the {\em entire} rotation group,
so these harmonic spaces are basically an enlargement of four-dimensional
space by the coset of the rotation group by H.
Conventional four-dimensional fields are recoverable from fields in such
harmonic spaces by performing an expansion in the harmonics on the
coset space. Such harmonic spaces are basically manifestly covariant versions
of
twistor spaces \cite{be}; and they can be used to construct
explicit local solutions by reformulating the Penrose-Ward twistor
transform in harmonic space  language. Moreover, just like twistor spaces,
they are amenable to supersymmetrisation.
Many four dimensional integrable systems have hitherto yielded themselves
to this harmonic-twistor method of describing the general solution:
the Yang-Mills  self-duality  equations  \cite{ym,goe}, all their
supersymmetric extensions
\cite{sym}, and the full N=3 super-Yang-Mills  equations  \cite{ber}.
Twistor theory, moreover, affords adaptation to curved spaces (reviewed in
e.g. \cite{pw}). Specifically, it yields a method of describing self-dual
solutions of Einstein equations (with or without a cosmological constant
\cite{ward2}). The harmonic space variant similarly allows itself to be
applied to the field equations describing hyper-K\"ahler \cite{GIOS,cmp}
and quaternionic spaces \cite{GIO}.

The harmonic-twistor method for self-dual  theories uses a presentation of
the equations in a harmonic space with $S^2$ harmonics as auxiliary
variables; and the essence of this version of the twistor transform is a
transformation to a system of coordinates, an `analytic' frame, in which
an invariant `analytic' subspace exists and in which the equations take the
form of Cauchy-Riemann-like (CR) equations. The method therefore encodes
the solution of non-linear equations  in certain `analytic' functions
(by which we mean functions depending only on coordinates of the invariant
`analytic' subspace). For instance, the Yang-Mills  self-duality  equations
 $F_{\mu\nu}={1\over 2} \epsilon_{\mu\nu\rho\sigma} F_{\rho\sigma}$
take the form (see e.g. \cite{ym,goe}) of the following system in harmonic
space with coordinates
$\{ x^{\pm \alpha}\equiv x^{\alpha a}u^\pm_a ,u^\pm_a\  ;
u^{+a}u^-_a = 1\ , u^\pm_a \sim e^{\pm\gamma} u^\pm_a \}$:
$$ [{\cal D}^+_\alpha, {\cal D}^+_\beta] = 0
= [{\cal D}^{++}, {\cal D}^+_\alpha]\   ,$$
where $ {\cal D}^+_\alpha
\equiv {\partial \over \partial {x^{-\alpha}}} + A^+_\alpha\   ,\quad
{\cal D}^{++} \equiv u^+_a {\partial \over \partial {u^-_a}}$.
Here $\alpha , a$ are 2-spinor indices and the signs
denote the conserved U(1) charge. The harmonics  $u^\pm_a$ are
fundamental SU(2) spinors; any representation of SU(2) allowing presentation
as a symmetrised product of them. Being defined up to a U(1)
transformation,
these harmonics parametrise $S^2 \simeq SU(2)/U(1) $. Moreover, since they
do so globally, these  $u^\pm_a$'s are much more convenient objects
than their Euler angle or stereographic parametrisations, allowing
the avoidance of the Riemann-Hilbert problem.
Having written the self-duality equations in the above harmonic space
language, the Frobenius argument allows the crucial
transformation to an `analytic' frame in which the covariant derivative
${\cal D}^+_\alpha$ takes the form of the flat derivative
${\partial \over \partial {x^{-\alpha}}}$,
completely trivialising the first commutation relation above; and in the
process shifting all the unsolved (dynamical) data to the harmonic
derivative, which loses its flatness by acquiring a connection:
${\cal D}^{++} \rightarrow
u^+_a {\partial \over \partial {u^{- a}}} + V^{++}$. This connection
$V^{++}$ then carries all the dynamical information and the equation of
motion for the connection $A^+_\alpha$ is replaced by the
Cauchy-Riemann-like analyticity condition:
$[{\cal D}^+_\alpha, {\cal D}^{++} ]
= {\partial \over \partial {x^{-\alpha}}} V^{++} = 0 $.
So the general local solution is encoded in an arbitrary analytic
$V^{++} = V^{++}(x^{+\alpha}, u^\pm_a)$. `Integrability' therefore becomes
manifest, though the problem of constructing specific explicit self-dual
vector potentials reduces to that of
inverting the above transformation for any specified analytic $V^{++}$.

This method has already been considered as a means of solving self-dual
gravity \cite{GIOS}, where the presence of vielbeins as well as connections
requires a suitable adaptation of the above flat-space Yang-Mills  strategy.
In this paper we reconsider this problem and show that even in the curved
case the essential features of the above strategy can be maintained.
In particular, we show that a special `half-flat' analytic coordinate frame
exists in which two of the four-dimensional covariant derivatives become
completely flat; and in which all the dynamics gets concentrated in the
covariant
harmonic derivative ${\cal D}^{++}$, just as in the Yang-Mills case, but
now in both the connection {\it and} vielbein parts of the latter. All these
parts of ${\cal D}^{++}$, moreover, can be solved for in terms of a
{\it single} arbitrary analytic prepotential, which therefore encodes the
general solution of the gravitational self-duality  conditions.
In the next section we discuss the harmonisation: the formulation of the
self-duality  conditions for the Riemann tensor in harmonic space.
Following \cite{GIOS} we then introduce (in section \ref{h-frame}) the class
of analytic coordinate frames which are distinguished by the property of
having an invariant analytic subspace and we discuss the corresponding
transformation rules in section \ref{transformations}.
We then list (in sect. \ref{cr}) the self-duality  equations,
`the CR system', for the vielbein and connection fields in these analytic
frames. This system has sufficient gauge freedom, affording the choice
(in section \ref{gauge}) of a particular analytic frame, the
`half-flat' gauge, in which a great
deal of the simplicity of the flat-space (Yang-Mills) construction
outlined above is recovered; and as a consequence (sect. \ref{solution})
the general local solution of the CR system (which in this gauge
takes a manifestly Cauchy-Riemann form) follows in
terms of a single arbitrary (i.e. unconstrained) analytic prepotential which
encapsulates the dynamics. Our refinement of the construction of \cite{GIOS}
considerably streamlines the procedure for the explicit construction of
self-dual metrics and corresponding spin connections. We describe this
procedure in section \ref{recipe};
and in section \ref{ex} we demonstrate it for the particular case of the
simplest nontrivial monomial choice of analytic prepotential which we
explicitly decode to reveal a metric related to the self-dual Taub-NUT
solution. In fact the form of the metric we obtain is
precisely that obtained by the alternative construction of hyper-K\"ahler
metrics using the harmonic superspace construction \cite{cmp} of N=2
supersymmetric sigma models, which have hyper-K\"ahler manifolds as target
spaces. In section \ref{concl} we discuss the relation to the alternative
Plebanski approach \cite{pleb} requiring solution of a second-order
differential equation. As a byproduct, our method yields a prescription
for the production of solutions to Plebanski's second `heavenly' equation,
though our construction is independent of this equation and does not
require its solution for the construction of self-dual  metrics.
In virtue of its generality, our method is a promising one for the explicit
construction of new local solutions to self-dual  gravity \cite{sdg}.
Further, it paves the way towards the solution of self-dual supergravity
theories and the explicit construction of supersymmetric hyper-K\"ahler
manifolds.

Our considerations are good for complexified space or for real spaces
of signature (4,0) or (2,2) (with appropriate handling of the latter as
a restriction of complexified space).
For concreteness however, we shall deal with the real Euclidean version,
with tangent space (structure) group being
the direct product $SU(2)\times SU(2)$; the first SU(2) having Greek spinor
indices $\alpha,\beta,...$, whereas we denote the second SU(2) by
Latin spinor indices $a,b,....$, ($\alpha, a = 1, 2$).
The covariant derivative ${\cal D}_{\alpha a}$
takes values in the tangent space algebra and defines the components
of the Riemann curvature tensor in virtue of the commutation relations
\cite{battelle}
$$[{\cal D}_{\beta b}, {\cal D}_{\alpha a}] =
\epsilon_{a b} R_{\alpha \beta} + \epsilon_{\alpha \beta} R_{a b},$$
with
$$\begin{array}{rl}
R_{\alpha \beta} &\equiv
C_{(\alpha\beta\gamma\delta)} \Gamma^{\gamma\delta}
 + R_{(\alpha\beta)(cd)} \Gamma^{cd} +{1\over 6}R \Gamma_{\alpha\beta} ,\cr
R_{a b} &\equiv
C_{(abcd)}\Gamma^{cd} + R_{(\gamma\delta)(ab)} \Gamma^{\gamma\delta}
+ {1\over 6} R \Gamma_{ab},\cr\end{array}$$
where round brackets denote symmetrisation and, in this spinor notation,
$C_{(abcd)} (C_{(\alpha\beta\gamma\delta)}) $ are the (anti-) self-dual
components of the Weyl tensor, $R_{(\alpha\beta) (c d)}$  are the components
of the tracefree Ricci tensor, $R$ is the scalar curvature,
$(\Gamma^{\gamma\delta},\Gamma^{cd})$ are generators
of the tangent space gauge algebra. We raise and lower all tangent space
indices using the antisymmetric invariant tensors $\epsilon^{\alpha\beta},
\epsilon^{ab}$ and $\epsilon_{\alpha\beta}, \epsilon_{ab}$ respectively,
e.g. $\epsilon^{ab}\psi_b = \psi^a ~,~ \epsilon_{ab} \psi^b = \psi_a ~;~
\epsilon_{12} =1.$
We presently take `self-duality' to  mean that the Riemann tensor is
self-dual, which in spinor notation means that $$R_{ab} =0 .$$
In virtue of the above (irreducible) decomposition this is tantamount to
the self-duality  of the Weyl tensor ($C_{(abcd)} = 0$) and the vanishing
of both the tracefree Ricci tensor and the scalar curvature
($R_{(\gamma\delta)(ab)} = R = 0$). These conditions clearly imply the
source-free Einstein Field Equations. In the present work we shall restrict
ourselves to this case of zero cosmological constant. Evidently, we may
write these self-duality  conditions in the form of the constraints
\begin{equation} [{\cal D}_{\beta b}, {\cal D}_{\alpha a}]
= \epsilon_{a b} R_{\alpha \beta}.\label{2}\end{equation}
Now, since $R_{(\gamma\delta)(ab)} = R = 0$, we have that
$R_{\alpha \beta} = C_{(\alpha\beta\gamma\delta)} \Gamma^{\gamma\delta}$.
The self-dual part of the curvature, $R_{\alpha \beta}$, therefore takes
values only in the SU(2) algebra with generators $\Gamma^{\gamma\delta}$.
This means that we may work  in a `self-dual  gauge' (see e.g. \cite{eh})
in which only this half of the tangent space group is localised: just the
$SU(2)$ labelled by indices $\alpha, \beta,....$, while the second $SU(2)$
(indices $a,b,...$) remains rigid. Correspondingly, the space coordinate
$x^{\mu a}$  only has one `world' spinor index  $\mu$, the second one
being identified with the tangent space spinor index $a$. Covariant
derivatives therefore take the form
\begin{equation} {\cal D}_{\alpha a} = E^{\mu b}_{\alpha a}\partial_{\mu b}
+ \omega_{a\alpha}\   ,\label{1}\end{equation}
where in this gauge the spin connections take values only in the  $SU(2)$
algebra (generators $\Gamma^\delta_\beta$) of the local structure group, viz.
\begin{equation} (\omega_{\alpha a})
= \omega_{\alpha a \sigma}^\delta (\Gamma^\sigma_\delta), \label{3}
\end{equation}
a restriction  which clearly implies that the curvatures also take values
only in this restriction of the tangent space algebra, i.e. are then
automatically self-dual. So with the connection in the form (\ref{3}),
eq.(\ref{2}) is no longer a constraint on the curvature; and the problem
reduces to that of finding vierbeins $E^{\mu b}_{\alpha a}$ satisfying
the conditions of zero torsion implicit in (\ref{2}). This is the principal
difference from the Yang-Mills  case, where there are neither vierbeins nor
torsions and the entire problem is that of solving the analogue of (\ref{2})
as curvature  constraints on the connection (as opposed to torsion
constraints on the vierbein).

In the above `self-dual  gauge', not only is the curvature automatically
self-dual, but since both connection and curvature are restricted to take
values in $SU(2)\simeq Sp(1)$, the metric is manifestly hyper-K\"ahler.
This way of considering  hyper-K\"ahler spaces, as solutions of the
self-duality conditions (\ref{2}), may immediately be generalised to higher
dimensions. Equations describing higher (4n) dimensional hyper-K\"ahler
spaces may be obtained by simply replacing the $SU(2)\simeq Sp(1)$
index $\alpha$ in (\ref{2}) by an $Sp(n)$ one \cite{GIOS}.
We shall presently restrict ourselves, however, to pure  self-dual gravity
in four dimensions, the treatment of $4n$-dimensional hyper-K\"ahler
manifolds for $n>1$ is evident (see sect. \ref{concl}).
\section{The self-duality  equations in harmonic space  }

Having an `ungauged' SU(2) part of the tangent space algebra at our disposal,
we `harmonise' it by introducing  $S^2$ harmonics
$\{ u^{\pm a}~ ;~ u^{+a}u^-_a =1 , u^\pm_a \sim e^{\pm\gamma} u^\pm_a \}$,
where a is an SU(2) spinor index and $\pm$ denote U(1) charges \cite{har}.
We begin by defining the covariant derivatives in
the {\em central} coordinate basis of {\it harmonic space  } thus:
\begin{equation} {\cal D}^\pm_\alpha \equiv
D^\pm_\alpha + \omega^\pm_\alpha = u^{\pm a} {\cal D}_{\alpha a},
\quad {\cal D}^{++}= \partial^{++} = u^+_a {\partial \over \partial {u^-_a}}
\label{4}\end{equation}
where the harmonic derivative ${\cal D}^{++}$ is a partial derivative acting
only on harmonics and is connection-less (this being the characteristic
feature of this basis).
\setcounter{equation}{0}
\renewcommand\theequation{5\alph{equation}}

{\it The following system in harmonic space is equivalent to the
self-duality conditions
(\ref{2}):}
\begin{eqnarray}
 [{\cal D}^+_\alpha, {\cal D}^+_\beta] &=&0 \\[0cm]
 [{\cal D}^{++}, {\cal D}^+_\alpha] &=& 0 \\[0cm]
 [{\cal D}^+_\alpha, {\cal D}^-_\beta] &=& 0
 \quad\hbox{(modulo  $R_{\alpha \beta}$)} \\[0cm]
[{\cal D}^{++}, {\cal D}^-_\alpha ] &=&{\cal D}^+_\alpha,
\end{eqnarray}
\setcounter{equation}{5}
\renewcommand\theequation{\arabic{equation}}
We obtain (5a) on multiplying (\ref{2}) by $u^{+a} u^{+ b}$.
Conversely, (5b) ensures linearity of ${\cal D}^+_\alpha$ in the harmonics,
so the harmonic (i.e. $u^+$-) expansion of (5a) yields (1) up to
possible torsion terms containing $\epsilon_{a b}$, viz.
 $\epsilon_{a b} T^{\gamma c}_{\alpha \beta} {\cal D}_{\gamma c} $.
(In the flat-space Yang-Mills case, recall, the relations (5a,b) {\em are}
equivalent to the self-duality conditions).  To exclude the existence of
such torsion terms we need to include the commutation relations (5c); and
(5d) is then required in order to ensure that in the present central basis
${\cal D}^-_\alpha$ contain harmonics only linearly (as in (\ref{4})).
The system of commutators (5a-d) is then equivalent to the original
self-duality relations (\ref{2})
\footnote{In this paper we shall deal only with these covariant
 derivatives $( {\cal D}^\pm_\alpha , {\cal D}^{++}  )$ which enter the
 system (5); we shall not use the additional harmonic covariant
 derivative ${\cal D}^{--}$ on which the discussion of \cite{GIOS} was
 based.}. The system (5) is in fact a Cauchy-Riemann-like (CR) system.
Only the coordinate frame needs to be changed in order to make its CR
nature manifest.
\section{The analytic frames}\label{h-frame}

The choice (\ref{4}) of covariant derivatives
corresponds to what we have called the {\it central frame}
with coordinates $\{x^{\mu \pm} \equiv x^{\mu a}u^\pm_a , u^\pm_a\}$.
The system of commutators (5), however, describes self-duality covariantly,
without reference to the particular form (\ref{4}) of ${\cal D}^{++}$ and
${\cal D}^\pm_\alpha$,
these covariant derivatives in general containing connections {\em and}
vielbeins, providing covariance under both SU(2) local frame transformations
and general coordinate transformations $\delta x^{\mu a} = \tau^{\mu a}(x)$.
In the above central coordinates the equivalence  $(5)\Leftrightarrow(1)$
is manifest, however this covariance may be exploited in order to
choose an alternative special coordinate system, the `half-flat' gauge
mentioned in the Introduction, in which the CR nature of (5) becomes
manifest instead. The latter coordinate system belongs to a select (gauge
equivalent) class of {\em analytic frames}, which have the distinguishing
feature of an invariant `analytic' subspace under general coordinate
transformations.
We call an object `analytic' if it is independent of a subset (viz.
$\{ x_h^{\mu -}\}$) of some new set of coordinates
$\{ x^{\mu \pm}_h, u^\pm_a \}$, with the
invariant `analytic' subspace having `holomorphic'
\footnote{This terminology, borrowed from complex analysis, is to be
understood only in this sense. We take our $x^{\mu a}, x_h^{\mu\pm}$
to be {\em real} coordinates. Appropriate hermiticity conditions for the
harmonics are discussed in the harmonic space  literature \cite{har,ym,goe}.
Of course, all coordinates can also be complexified (see \cite{goe}).}
coordinates $ x_h^{\mu+}\;$.
Any such new coordinates $x^{\mu\pm}_h$ are of course
some nonlinear functions of the central frame
(or customary  $x^{\mu a}$) coordinates and of the harmonics $u^\pm_a$. :
\begin{equation} x^{\mu a} \rightarrow  x^{\mu \pm}_h
=  x^{\mu \pm}_h (x^{\mu a}u^\pm_a , u^\pm_a)\  ,\label{6}\end{equation}
The necessity of determining these h-coordinates as (invertible) functions
of the central frame ones is the main novel feature of the curved-space
construction and this is the crucial difference from the flat self-dual
Yang-Mills case,
where we have simply the linear relationship of the central coordinates
$x^{\mu \pm} = x^{\mu a}u^\pm_a$.

In order to have an invariant `analytic' subspace, the functions
$x^{\mu \pm}_h (x^{\mu a}u^\pm_a , u^\pm_a)$ in (\ref{6}) are clearly
required to have the crucial property that under the
mapping (\ref{6}) the covariant derivatives $D^+_\alpha$ in (\ref{4})
contain derivations with respect to $x_h^{\mu -}$ only. In general,
(\ref{6}) induces the mapping:
$$ D^+_\alpha \rightarrow (D^+_\alpha x^{\mu -}_h) \partial^+_{h\mu}
+ (D^+_\alpha x^{\mu +}_h) \partial^-_{h\mu},$$
where $\partial^+_{h \mu} = {\partial \over \partial {x^{\mu -}_h}}$,
$\partial^-_{h \mu} = {\partial \over \partial {x^{\mu +}_h}}$,
so the requirement that only  $\partial^+_{h\mu}$-terms appear on the right
is tantamount to the condition that the holomorphic coordinates
preserve the flat space relation
\begin{equation} D^+_\alpha x^{\mu +}_h = 0 .\label{7}\end{equation}
We take this to be the {\it defining condition} for analytic frame
coordinates. This yields
\begin{equation} D^+_\alpha = (D^+_\alpha x^{\mu -}_h) \partial^+_{h\mu}
\equiv f^\mu_\alpha  \partial^+_{h\mu} ,\label{8}\end{equation}
where we have defined an analytic frame zweibein $f^\mu_\alpha$. Having only
derivatives with respect to $x^{\mu -}$  on the right, the conditions
$D^+_\alpha \Psi = 0$ then imply (for invertible zweibein $f^\mu_\alpha$)
the required `analyticity' of $\Psi$,
i.e. the independence of $x_h^{\mu -}$, $\partial^+_{h\mu}\Psi = 0$.

On the other hand, the negatively charged derivatives $D^-_\alpha$ in
(\ref{4}) contain derivations with respect to all the new coordinates:
\begin{equation} D^-_\alpha
= - e^\mu_\alpha \partial^-_{h \mu} + e^{--\mu}_\alpha \partial^+_{h \mu} ,
\label{9}\end{equation}
where we have defined a further neutral zweibein,
\begin{equation} e^\mu_\alpha
= -D^-_\alpha x^{\mu +}_h, \label{10}\end{equation}
(the minus sign is chosen so as to have $e^\mu_\alpha=\delta^\mu_\alpha$ in
the flat-space limit) and a doubly-negatively charged one
\begin{equation} e^{--\mu}_\alpha
= D^-_\alpha x^{\mu -}_h .\label{11}\end{equation}

We now come to the harmonic derivative. Being flat in the central
coordinates, ${\cal D}^{++} = \partial^{++}$, it acquires vielbeins in
the holomorphic ones:
\begin{equation} \partial^{++} \rightarrow \Delta^{++}
=\partial^{++} + H^{++ \mu +}\partial^-_{h \mu}
  + (x^{\mu +}_h + H^{++ \mu -}) \partial^+_{h\mu}.\label{12}\end{equation}
where the vielbeins are defined in terms of the holomorphic coordinates thus:
\begin{equation} H^{++\mu+}
 = \partial^{++} x^{\mu +}_h               ,\label{13}\end{equation}
\begin{equation} H^{++ \mu -}
 =\partial^{++} x^{\mu -}_h - x^{\mu +}_h.\label{14}\end{equation}
Note that these vielbeins are defined so as to have
$H^{++\mu +} = H^{++\mu -} = 0$ in the flat-space limit.

Now, the curvature in (5a) being zero, we may perform an SU(2) structure
group transformation making the covariant derivatives ${\cal D}^+_\alpha$
connectionless, as in the Yang-Mills case discussed in the introduction.
Namely, (5a) implies the existence of an invertible matrix
$\varphi^{\breve\beta}_\beta $ (having inverse
$(\varphi^{-1} )_{\breve\beta}^\beta\; ; \;
(\varphi^{-1} )^{\breve\beta}_\beta \varphi^\beta_{\breve\alpha}
= \delta^{\breve\beta}_{\breve\alpha}\; , \;
\varphi^{\breve\beta}_\beta (\varphi^{-1} )_{\breve\beta}^\alpha
= \delta^\alpha_\beta \;$)  satisfying the equations
\begin{equation} {\cal D}^+_\alpha \varphi \equiv
(D^+_\alpha + \omega^+_\alpha) \varphi = 0 \label{15}\end{equation}
which imply the pure-gauge form of the connection in terms of $\varphi$ :
\begin{equation} ({\cal D}^+_\alpha)^\gamma_\beta =  D^+_\alpha
\delta^\gamma_\beta - D^+_\alpha \varphi_\beta^{\breve\beta}
\varphi^{-1\gamma}_{\breve\beta} .\label{16}\end{equation}

We may therefore use a solution of (\ref{15}) to perform an SU(2)
transformation in order to `gauge away' the connection in
${\cal D}^+_\alpha$:
$$ {\cal D}^+_\alpha \rightarrow  \varphi^{-1} {\cal D}^+_\alpha \varphi
= D^+_\alpha $$  with $D^+_a $ as in (\ref{8}) above. Now, under this
transformation, just as in the Yang-Mills case, the harmonic derivative
${\cal D}^{++}$  acquires a connection:
\begin{equation}
D^{++} \rightarrow {\cal D}^{++} = \varphi^{-1} [D^{++}] \varphi
\equiv \Delta^{++} + \omega^{++}, \label{17}\end{equation}
where \begin{equation}
\omega^{++} = \varphi^{-1} \Delta^{++} \varphi .\label{18}\end{equation}
We shall show that in a particular analytic frame the equations implied
by the CR system (5) for the analytic frame vielbeins and connections
defined above may be solved (treating the h-coordinates
$\{ x^{\mu \pm}_h, u^\pm_a \}$ as independent variables)
in terms of an arbitrary analytic prepotential.
Our strategy will then be to solve equations (\ref{13},\ref{14},\ref{18})
for $x_h^{\mu+},x_h^{\mu-}$, and $\varphi$ respectively, for some specific
choice of analytic prepotential (treating, in turn, the central coordinates
$\{x^{\mu a}, u^\pm_a \}$ as independent variables).
Having determined the latter data, we shall invert of the mapping (\ref{6})
and obtain the vierbein in (\ref{1}) explicitly. The self-dual  metric
will then afford immediate construction. The problem of the explicit
construction will therefore reduce to that of solving
eqs.(\ref{13},\ref{14},\ref{18}) for fields $H^{++\mu\pm}$ and
$\omega^{++}$ determined by some specified choice of the analytic
prepotential.
\section{ Transformation rules}\label{transformations}

There exist two kinds of the tangent-space transformations; central frame
ones with local parameters $\tau^\beta_\alpha(x^{\mu a})$ and analytic
frame ones with local parameters
$\lambda^{\breve\alpha}_{\breve\beta}(x_h^{\mu+}, u)$ (we distinguish
$\lambda$-transforming indices by a `breve' accent).
The matrix $\varphi$ is defined up to the local gauge equivalence
\begin{equation} \varphi^{{\breve\alpha}}_\alpha \sim
\tau_\alpha^\beta(x^{\mu a}) \varphi_\beta^{\breve\beta}
\lambda_{\breve\beta}^{\breve\alpha}(x_h^{\mu+}, u) \label{19}\end{equation}
The connection in (\ref{15}) does not transform under the analytic
transformations parametrised by $\lambda$, these being `pregauge' in the
central frame; whereas
for the analytic frame connection $\omega^{++}$ in (\ref{18}) the tangential
transformations parametrised by $\tau$ are `pregauge' and therefore
leave this connection {\em invariant}.
Consider some spinor $F_\alpha$, which has tangent transformation
$$\delta F_\alpha = \tau^\beta_\alpha(x^{\mu a}) F_\beta.$$ The parameters
$\tau^\beta_\alpha(x)$ being
non-analytic, this transformation would not be consistent if $F_\alpha$
were analytic; i.e. the $\tau$-transformation does not preserve the
analytic subspace. However using $\varphi$
we can pass to a spinor having breved indices,
$F_{\breve\alpha} = (\varphi^{-1})_{\breve\alpha}^\beta F_\beta ,\quad
F_\alpha = \varphi_\alpha^{\breve\beta} F_{\breve\beta}$
which has tangent transformations preserving the analytic subspace:
$$\delta F_{\breve\alpha}
= \lambda_{\breve\alpha}^{\breve\beta}(x^+_h, u) F_{\breve\beta}.$$
So  $\varphi$ is clearly a bridge taking us from central frame tangent-space
indices ($\alpha$) to analytic frame tangent-space ones (${\breve\alpha}$).
In the analytic frame
we shall use only quantities with breved tangent-space spinor indices, using
as many $\varphi$'s as are necessary in order to obtain the suitably
transforming quantity. In the  CR system (5) therefore, we pass to
appropriately transforming covariant derivatives:
$$ {\cal D}^+_{\breve\alpha} = (\varphi^{-1})_{\breve\alpha}^\beta
{\cal D}^+_\beta \quad {\cal D}^-_{\breve\alpha} =
(\varphi^{-1})_{\breve\alpha}^\beta {\cal D}^-_\beta. $$

In  gravitation theory the most important transformations are the world
ones. For the central frame coordinates these form the diffeomorphism group
with local parameters $\tau^{\mu a}(x)$, i.e. $\delta x^{\mu a}=
\tau^{\mu a}(x)$. In an analytic frame, by definition (see sect.III),
harmonic space diffeomorphisms preserve analyticity, i.e. leave the
analytic subspace (the analytic planes with coordinates $x_h^{\mu+}$)
of harmonic space invariant. Namely,
\begin{equation} \delta x^{\mu +}_h =\lambda^{\mu +}(x^+_h, u)
,\label{21}\end{equation} whereas
\begin{equation} \delta x^{\mu -}_h = \lambda^{\mu -}(x^+_h, x^-_h, u)
 .\label{22}\end{equation}
The important property being of course that {\it the positively  charged
parameters
are analytic, whereas the negatively charged ones are not}, implying
that the most general diffeomorphism preserves the analytic subspace.
The harmonics also allow transformation, requiring consideration
of the complexified picture \cite{goe}. However we do not need to consider
these transformations since they do not affect the CR system (5).
Such transformations, however, are necessary in the case of non-zero
cosmological constant (see \cite{GIO}).

{}From the covariance of the covariant derivatives
(\ref{8},\ref{9},\ref{12}) under the transformations (\ref{21},\ref{22}),
we obtain the following transformation rules for the vielbeins introduced
in section \ref{h-frame}:
 \begin{equation}  \delta f^\mu_{\breve\alpha}
 = f^\nu_{\breve\alpha} \partial^+_{h \nu}\lambda^{\mu -} +
  \lambda^{\breve\beta}_{\breve\alpha} f^\mu_{\breve\beta} ,\label{23}
  \end{equation}
  \begin{equation}  \delta e^\mu_{\breve\alpha}
  = e^\nu_{\breve\alpha} \partial^-_{h \nu} \lambda^{\mu +}
 +  \lambda^{\breve\beta}_{\breve\alpha} e^\mu_{\breve\beta} ,\label{24}
 \end{equation}
\begin{equation} \delta e^{--\mu}_{\breve\alpha}
= - e^\nu_{\breve\alpha} \partial^-_{h \nu}\lambda^{\mu -}
+ e^{--\nu}_{\breve\alpha} \partial^+_{h \nu}\lambda^{\mu -}
+ \lambda^{\breve\beta}_{\breve\alpha} e^{--\mu}_{\breve\beta} ,\label{25}
\end{equation}
\begin{equation} \delta H^{++\mu +}
= \Delta^{++} \lambda^{\mu +},\label{26}\end{equation}
\begin{equation} \delta H^{++ \mu -}
= \Delta^{++} \lambda^{\mu -} - \lambda^{\mu +}.\label{27}\end{equation}
\section{ The Cauchy-Riemann system of equations}.\label{cr}

We now examine the content of the CR system (5) in an analytic frame,
i.e. with the covariant derivatives taking the explicit form
$$\begin{array}{rl} {\cal D}^+_{\breve\alpha}
=& f^\mu_{\breve\alpha} \partial^+_{h\mu}  \cr
{\cal D}^-_{\breve\alpha} =& -e^\mu_{\breve\alpha} \partial^-_{h\mu}
+ e^{--\mu}_{\breve\alpha} \partial^+_{h\mu}  + \omega^-_{\breve\alpha} \cr
{\cal D}^{++}  =& \partial^{++} + H^{++ \mu +}\partial^-_{h \mu}
           + (x^{\mu +}_h + H^{++ \mu -}) \partial^+_{h\mu}  + \omega^{++}\
           .\end{array}$$
Not all the equations implied by (5) for the vielbein and connection
fields in these covariant derivatives are `dynamical' in character,
in the sense of requiring solution for the determination of the metric.
We shall first extract the set of such `dynamical' equations,
the remaining equations basically determining redundant fields.

For the zweibein $f^\mu_{{\breve\beta}}$  we have from (5a) the equations
\begin{equation}
f^\nu_{[{\breve\alpha}} \partial^+_{h \nu} f^\mu_{{\breve\beta}]} = 0.
\label{28}\end{equation}

The vanishing of the torsion coefficients of  $\partial^-_{h\mu}$ in (5c)
and (5b) requires the vielbeins $e^\mu_{\breve\alpha}$ and $H^{++ \mu +}$,
respectively, to be {\it analytic}:
\begin{equation}
D^+_{\breve\alpha} e^\mu_{\breve\beta} = 0\      ,\label{29}\end{equation}
\begin{equation}
D^+_{\breve\alpha} H^{++ \mu +} = 0\   .\label{30}\end{equation}
The vanishing of the curvature in (5b) yields a further analyticity
condition; for the connection $\omega^{++}$,
\begin{equation}
D^+_{\breve\alpha} \omega^{++} = 0\       .\label{31}\end{equation}
The solution of this equation is however not independent of the solution of
the previous two analyticity conditions; the equation
\begin{equation} -{\cal D}^{++} e^\mu_{\breve\alpha}
    - {\cal D}^-_{\breve\alpha} H^{++\mu+} = 0 ,\label{32}\end{equation}
which is a consequence of the vanishing of the torsion coefficients of
$\partial^-_{h \mu}$ in (5d), provides an important constraint amongst the
three analytic fields $e^\mu_{\breve\beta}\ , H^{++\mu+}$, and $\omega^{++}$.
Further, these fields determine  $H^{++\mu-}$ in virtue of the equation
\begin{equation}   D^+_{\breve\alpha} H^{++ \mu -}
    = {\cal D}^{++} f^\mu_{\breve\alpha}  \   ,\label{33}\end{equation}
which arises from the requirement of the vanishing of the torsion
coefficients of $\partial^+_{h\mu}$  in constraint (5b).

In order to solve self-dual  gravity (\ref{2}), it suffices to solve the
above set of equations (\ref{28}-\ref{33}). The remaining equations from (5)
are basically conditions determining consistent expressions for the fields
$e^{--\mu}_{\breve\beta}$ and $\omega^-_{\breve\alpha}$, whose determination
is actually not necessary in order to find self-dual  metrics. These fields
represent the same degrees of freedom as the fields
$\{f^\mu_{\breve\alpha}, e^\mu_{\breve\alpha}, H^{++\mu\pm}, \omega^{++} \}$
and therefore represent redundant degrees of freedom. The field
$e^{--\mu}_{\breve\alpha}$ is determined by the equation following from
the equality of the coefficients of $\partial^+_{h \mu}$ in (5d), namely,
\begin{equation} {\cal D}^{++} e^{--\mu}_{\breve\alpha} =
  f^\mu_{\breve\alpha}
    + {\cal D}^-_{\breve\alpha} (H^{++ \mu -} + x^{\mu +}_h)
     .\label{34}\end{equation}
The vanishing of torsion coefficients of $\partial^+_{h \mu}$ in (5c) yields
\begin{equation}  D^+_{\breve\alpha} e^{--\mu}_{\breve\beta}
= {\cal D}^-_{\breve\beta} f^\mu_{\breve\alpha} \  ,\label{35}\end{equation}
which together with the condition obtained from the requirement that
the antisymmetric part of the curvature in (5c) vanishes, i.e.
 \begin{equation}  D^+_{[{\breve\alpha}} \omega^-_{{\breve\beta}]} = 0,
 \label{36}\end{equation}
determines $\omega^-_{\breve\beta}$, which satisfies the final equation
contained in (5), namely the vanishing of the curvature in (5d),
\begin{equation} {\cal D}^{++} \omega^-_{\breve\alpha}
      - {\cal D}^-_{\breve\alpha} \omega^{++}  = 0 ,\label{37}\end{equation}
automatically, in virtue of (\ref{34}).
\section{The `half-flat' gauge}\label{gauge}

The set of fields satisfying the system of equations listed in the
previous section is actually highly redundant, possessing the gauge
invariances (\ref{23}-\ref{27}). The system may therefore be reduced
by fixing the parameters $\lambda^{\breve\beta}_{\breve\alpha}(x^+_h ,u),
\lambda^{\mu+}(x^+_h ,u)$, and $\lambda^{\mu-}(x^\pm_h ,u)$,
and thereby specifying local coordinates. Remarkably, in a suitable
coordinate gauge, this system of equations becomes manifestly soluble.
Firstly, since  $e^\mu_{\breve\alpha}$ is analytic (\ref{29}), the gauge
invariance (\ref{23}) with analytic parameters
$\lambda^{\breve\beta}_{\breve\alpha}$, allows us to
choose local coordinates $x^{\mu+}_h$ such that $e^\mu_{\breve\alpha}$ is
a unit matrix. Further, the torsion constraint (5a) essentially says,
by Frobenius' integrability theorem, that ${\cal D}^+_{\breve\alpha}$ is
gauge-equivalent to the partial derivative $\partial^+_{\breve\alpha}$.
There therefore exists a coordinate gauge in which (\ref{28}) is solved
in virtue of the zweibein $f^\mu_{\breve\alpha}$ also taking the form
of a unit matrix.
Choosing such a gauge in which ${\cal D}^+_{\breve\alpha}$ is completely
flat is therefore tantamount to changing coordinates (using gauge degrees
of freedom parameterised by $\lambda^{\mu-}$ (\ref{24})) thus:
$x_h^{\mu -} \rightarrow y^{\mu -} =
y^{\mu -} (x_h^{\nu +},x_h^{\nu -}), \quad
x_h^{\mu +} \rightarrow y^{\mu +} =
y^{\mu +} (x_h^{\nu +})$, such that in the old coordinates
\begin{equation}  f^\mu_{\breve\alpha} =
{\partial { x^{\mu -}_h}\over \partial {y^{{\breve\alpha}-}}}\  ,
\label{frob}\end{equation} which manifestly solves (\ref{28}) since
$${\partial x^{\nu -}_h \over
\partial y^{[{\breve\alpha} -}} {\partial \over \partial {x^{\nu -}_h}}
{\partial \over \partial {y^{{\breve\beta}] -}}} x^{\mu -}_h
= {\partial \over \partial {y^{[{\breve\alpha} -}}}
{\partial \over \partial {y^{{\breve\beta}] -}}} x^{\mu -}_h \equiv 0. $$
With this zweibein ${\cal D}^+_{\breve\alpha}$ is clearly flat in the
new coordinates:
$${\cal D}^+_{\breve\alpha}
= {\partial \over \partial {y^{{\breve\alpha} -}}} ,$$
the constraint (5a) becoming an identity and Frobenius' theorem
becoming manifest.

\begin{equation} f^\mu_{\breve\alpha} = \delta^\mu_{\breve\alpha}, \quad
e^\mu_{\breve\alpha} = \delta^\mu_{\breve\alpha}\   .\label{h-flat}
\end{equation}
In this special `half-flat' gauge the distinction between world and
tangent indices has evidently been eliminated and only the set of vielbein
and connection fields
$\{ H^{++\mu\pm}, e^{--\mu}_{\breve\alpha} , \omega^{++} ,
\omega^-_{\breve\alpha} \}$ remain, of which, as we shall see, those
contained in ${\cal D}^{++}$, namely
$\{ H^{++\mu\pm}, \omega^{++} \}$ contain all the dynamical information.
This Yang-Mills-like feature is the distinguishing one of this particular
analytic frame which we shall henceforth use, although we shall continue
to call the coordinates in this particular gauge $x_h^{\mu \pm}$ rather
than $y^{{\breve\alpha} \pm}$.

In this gauge, residual gauge transformations have
parameters constrained by relations from (\ref{23},\ref{24}), viz.
$$ \partial^-_{h {\breve\alpha}}\lambda^{\mu +} + \lambda^\mu_{\breve\alpha}  =
0, \quad
\partial^+_{h {\breve\alpha}}\lambda^{\mu -} + \lambda^\mu_{\breve\alpha}  =
0 .$$ So the residual diffeomorphism parameters
$\lambda^{\mu \pm}(x_h^+, u)$
are no longer arbitrary but are constrained by the relations
$$ \partial^-_{h \mu}\lambda^{\mu +} = 0, \quad
\partial^+_{h \mu}\lambda^{\mu -} = 0\  ,$$
since the tangent parameters $\lambda^\mu_{\breve\alpha}$ are traceless.
It follows that the thus constrained $ \lambda^{\mu +}$ can be expressed
in terms of an unconstrained
doubly charged {\it analytic} parameter $\lambda^{++}$:
\setcounter{equation}{0}
\renewcommand\theequation{39\alph{equation}}
\begin{equation} \lambda^{\mu +}_{{\rm res}}(x_h^+, u) =
\partial^{\mu -}_h \lambda^{++}(x_h^+, u)\   .\end{equation}
These diffeomorphism parameters in turn determine the Lorentz ones, the
residual tangent transformations actually being induced by the world ones:
\begin{equation} (\lambda^\mu_{\breve\alpha})_{{\rm res}} =
-\partial^-_{h{\breve\alpha}}\lambda^{\mu +}_{{\rm res}}(x_h^+, u)\  .
\end{equation}
As for the remaining $\lambda^{\mu -}$ transformations, these have
parameters:
\begin{equation}  (\lambda^{\mu -})_{{\rm res}} =
\partial^-_{h \nu} \lambda^{\mu +}(x_h^+, u) x_h^{\nu -} +
\tilde\lambda^{\mu -}(x_h^+, u)\   ,\end{equation}\setcounter{equation}{39}
\renewcommand\theequation{\arabic{equation}}
\noindent where $ \tilde\lambda^{\mu -}(x_h^+, u)$ is an arbitrary
{\it analytic} parameter \footnote {When considering self-dual  gravity
in \cite{GIOS} the alternative gauge  condition
$$\Delta^{++} \lambda^{\mu -} = \lambda^{\mu +} $$ corresponding to the
preservation of the flat space relation  $\partial^{++} x^{\mu -} =
x^{\mu -}$ was adopted.  This condition completely fixes the
gauge parameters $\lambda^{\mu -}$. We presently choose to avoid this
lack of freedom, preferring the more convenient gauge (\ref{h-flat}).}.
The remaining vielbeins $H^{++\mu+}, H^{++\mu -}$, and
$e^{--\mu}_{\breve\alpha}$ still
transform according to  (\ref{26},\ref{27},\ref{25}), respectively, with
parameters being the residual ones (39).

\section{ The analytic frame solution} \label{solution}

We shall now show that in the `half-flat' gauge (\ref{h-flat}) with
covariant derivatives taking the form
\begin{eqnarray} {\cal D}^+_{\breve\alpha}
&=&  \partial^+_{h{\breve\alpha}}  \cr
{\cal D}^-_{\breve\alpha}
&=& - \partial^-_{h{\breve\alpha}}
+ e^{--\mu}_{\breve\alpha} \partial^+_{h\mu}  + \omega^-_{\breve\alpha}
\label{h-deriv}\\[0pt]
{\cal D}^{++}  &=& \partial^{++} + H^{++ \mu +}\partial^-_{h \mu}
           + (x^{\mu +}_h + H^{++ \mu -}) \partial^+_{h\mu}  + \omega^{++}\
           ,\nonumber\end{eqnarray}
the system of equations (\ref{28},\ref{29}), or equivalently the self-duality
system (5) becomes manifestly soluble. Moreover, our main claim is that:

{\it An unconstrained analytic prepotential, ${\cal L}^{+4}$, encodes all
local information on self-dual gravity.}

To prove this claim we begin by recalling that in the `half-flat' gauge
the difference between
world and tangent indices has become rather conventional, all essential
information about the manifold having moved to the vielbeins and
connections in ${\cal D}^{++} $. Eqs.(\ref{28},\ref{29}) clearly drop out in
this gauge, leaving, from the dynamical set (\ref{28}-\ref{33}), only
the analyticity conditions for
$H^{++\mu+}$ and $\omega^{++}$ (\ref{30},\ref{31}) and the relationships
(\ref{32}) and (\ref{33}). These four equations, and therefore the analytic
frame self-duality  conditions, can be consistently solved
in terms of a single arbitrary analytic prepotential of charge $+4$.
To prove that such a prepotential exists (at least locally) we begin with an
arbitrary {\it analytic} $H^{++{\breve\beta} +}$ (satisfying (\ref{30})).
{}The relation (\ref{32}) then yields an expression for the harmonic
connection which is manifestly analytic, automatically satisfying its
equation of motion (\ref{31}),
\begin{equation}  \omega^{++{\breve\beta}}_{\breve\alpha} =
\partial^-_{h {\breve\alpha}} H^{++{\breve\beta} +} .\label{40}\end{equation}
Now the requirement of tracelessness of this connection yields a constraint
on $H^{++{\breve\beta} +}$ which has local solution in terms of the sought
unconstrained analytic prepotential  ${\cal L}^{+4}$, i.e.
\begin{equation}  H^{++\mu +}  = \partial^{-\mu}_h {\cal L}^{+4} =
\epsilon^{\mu\nu} {\partial{{\cal L}^{+4}} \over \partial {x^{+\nu}_h}}\  .
\label{41}\end{equation}
The transformation rule (\ref{26}) for $H^{++\mu +}$ induces the
gauge transformation
\begin{equation} \delta {\cal L}^{+4} = \Delta^{++}\lambda^{++} +
\partial^{\mu -}_h \lambda^{++} \partial^-_{h \mu} {\cal L}^{+4} =
\partial^{++}\lambda^{++},  \label{freedom}\end{equation}
where $\lambda^{++}$ is the unconstrained analytic gauge parameter in (39a).
So prepotentials differing by the harmonic partial derivative of an analytic
function correspond to gauge equivalent solutions of the CR system.
Eq.(\ref{33}) remains
and yields a relationship, using (\ref{40}), between the two vielbeins
in ${\cal D}^{++} $:
\begin{equation} \partial^+_{h {\breve\alpha}} H^{++ {\breve\beta} -} =
\omega^{++{\breve\beta}}_{\breve\alpha} = \partial^-_{h {\breve\alpha}}
H^{++{\breve\beta} +}\  .\label{42}\end{equation}
Integrating this equation, we obtain
\begin{equation}  H^{++ {\breve\beta} -} =
x_h^{{\breve\alpha} -} \partial^-_{h {\breve\alpha}}  H^{++{\breve\beta} +}
= x_h^{{\breve\alpha} -} \partial^-_{h {\breve\alpha}}
\partial^{-{\breve\beta}}_h {\cal L}^{+4} ,\label{43}\end{equation}
up to an arbitrary analytic function absorbed by the gauge freedom (\ref{27}).

We can therefore determine all the required fields ($H^{++\mu\pm}$ and
$\omega^{++}$) consistently, i.e. solve the dynamical content of (5), in
terms of the an unconstrained (i.e. arbitrary) analytic prepotential
${\cal L}^{+4}$. For the sake of completeness we show in appendix \ref{A}
that all the other equations from (5) are also indeed solved in terms of
${\cal L}^{+4}$ and determine the remaining analytic frame fields
($e^{--\mu}_{\breve\alpha} , \omega^-_{\breve\alpha}$) as
functionals of ${\cal L}^{+4}$.

The conventional vierbeins and connections in the central basis may now be
constructed according to the procedure we give in the next section.
The correspondence between self-dual  metrics and prepotentials
${\cal L}^{+4}$ is, however, not unique, since prepotentials related by the
gauge transformation (\ref{freedom}) correspond to equivalent metrics.
We may fix this freedom by using the {\em normal gauge} \cite{GIOS}
in which ${\cal L}^{+4}$ has no explicit dependence on $u^+_a$ i.e.
${\cal L}^{+4} = {\cal L}^{+4}(x_h^{\mu +}, u^-_a)$. In other words any
explicit $u^+$-dependence may always be gauged away using the freedom
(\ref{freedom}). Consider the harmonic expansion of an arbitrary
prepotential having explicit $u^+$-dependence,
\begin{equation}
 {\cal L}^{+4}(x_h^+,u^+, u^-) = {\cal L}_{normal}^{+4}(x_h^+, u^-)
  + \sum_{n,m}  u^{+ a_1}\dots u^{+ a_n} u^{- b_1}\dots u^{- b_m}
                f^{4-n+m}_{(a_1 \dots a_n b_1 \dots b_m)}(x_h^+)\ ,
                \end{equation}
where the coefficients $f^{4-n+m}_{(a_1 \dots a_n b_1 \dots b_m)}(x_h^+)$
are monomials of degree $(4-n+m)$ in $x_h^{1+},\  x_h^{2+}$.
Now every term in the sum on the right may easily be shown to be a harmonic
partial derivative of an analytic function, so the entire sum has the form
$\partial^{++} \lambda^{++}(x_h^+,u^+, u^-)$, which may be absorbed by the
gauge freedom (\ref{freedom}), yielding the normal gauge form of
${\cal L}^{+4}$.

\section{ The reconstruction of vierbeins and connections in the
central frame} \label{recipe}

As we have seen, the analytic prepotential ${\cal L}^{+4}$ encodes all the
analytic basis dynamical information. But how does one extract the self-dual
metric in the original central basis from it? We now outline the procedure
for doing this starting from some specified analytic prepotential
${\cal L}^{+4}$.

A. From (\ref{41}) and (\ref{43}) obtain the vielbeins of ${\cal D}^{++}$:
$$\begin{array}{lcr} H^{++\mu+} &=& \partial^{-\mu}_h {\cal L}^{+4} \cr
H^{++\mu-} &=& x_h^{{\breve\alpha} -} \partial^-_{h {\breve\alpha}}
                \partial^{-\mu}_h {\cal L}^{+4} \end{array} $$

B. Consider (\ref{13}) as equations for the holomorphic coordinates
$x^{\mu +}_h$:
\begin{equation} \partial^{++} x^{\mu +}_h = \partial^{-\mu}_h {\cal L}^{+4}
                   .\end{equation}
Integrating these first order equations find $x^{\mu +}_h$ as functions
of the {\it central} frame coordinates $x^{\mu \pm}
(\equiv x^{\mu a} u^\pm_a )$ and the harmonics.

C. Having obtained $x^{\mu +}_h$, similarly solve (\ref{14}),i.e.
\begin{equation}
\partial^{++} x^{\mu -}_h = x^{\mu +}_h + x_h^{{\breve\alpha} -}
       \partial^-_{h {\breve\alpha}} \partial^{-\mu}_h {\cal L}^{+4}
\end{equation}
in order to determine
$x^{\mu -}_h$ as a function of the central frame coordinates.

D. From (\ref{40}) obtain the connection of ${\cal D}^{++} $:
\begin{equation} \omega^{++{\breve\beta}}_{\breve\alpha} =
\partial^-_{h {\breve\alpha}} \partial^{-{\breve\beta}}_h {\cal L}^{+4}
  ,\end{equation}
and using the results of steps B and C, express it explicitly
in terms of central coordinates.

E. With the $\omega^{++}$ obtained in step D, solve equation (\ref{18})
rewritten in the central frame
$\partial^{++} \varphi = \varphi \omega^{++}$, i.e.
\begin{equation}
\partial^{++} \varphi_\alpha^{\breve\beta} = \varphi_\alpha^{\breve\alpha}
\partial^-_{h {\breve\alpha}} \partial^{-{\breve\beta}}_h {\cal L}^{+4}
\label{44}\end{equation}
in order to obtain the bridge $\varphi$ in central coordinates.

F. Using results of steps B and C evaluate the partial derivatives
${\partial{ x^{\nu \pm}} \over \partial {x^{\mu -}_h}}$.

\noindent  The above data affords the immediate construction of
explicit self-dual  vierbeins, metrics, and connections as follows:

G. Multiply the bridge $\varphi$ obtained in step E with one of the
matrices of coordinate differentials from step F, and extract
the self-dual  vierbein from the relation
\begin{equation}  \varphi^\nu_\alpha
{\partial {x^{\mu -}}\over \partial {x^{\nu -}_h}}
= u^{+a} E^{\mu b}_{\alpha a}u^-_b\   ,\label{46}\end{equation}
using the completeness relation
$ u^{+a} u_b^{-} - u^{-a} u_b^{+} = \delta^a_b $.
Invert this vierbein and obtain the self-dual  metric
\begin{equation}  ds^2 = \epsilon_{ac}\epsilon_{\alpha\beta}
E^{\alpha a}_{\mu b} E^{\beta c}_{\nu d} dx^{\mu b}dx^{\nu d}\
                 .\label{48}\end{equation}

The proof of the relation (\ref{46}) is as follows. The central frame
vierbeins are given by the equation $$
(\varphi^{-1})^\alpha_{\breve\alpha} u^{+a} E^{\mu b}_{\alpha a}
{\partial{x^{\nu -}_h} \over \partial {x^{\mu b}}} = f^\nu_{\breve\alpha}$$
which follows from  equations (\ref{1},\ref{4},\ref{8}, and \ref{17}).
Fixing the gauge (\ref{h-flat}) and introducing quantities
\begin{equation}  Z^\mu_\alpha = u^{+a} E^{\mu b}_{\alpha a}u^-_b, \quad
Z^{\mu ++}_\alpha =u^{+a} E^{\mu b}_{\alpha a}u^+_b \label{45}\end{equation}
we obtain the system of equations
$$ Z^\mu_\alpha {\partial{x^{\nu -}_h} \over \partial {x^{\mu -}}} +
Z^{\mu ++}_\alpha {\partial{x^{\nu -}_h} \over \partial {x^{\mu +}}} =
\varphi^\nu_\alpha, $$
$$ Z^\mu_\alpha   {\partial{x^{\nu +}_h} \over \partial {x^{\mu -}}} +
Z^{\mu ++}_\alpha {\partial{x^{\nu +}_h} \over \partial {x^{\mu +}}} = 0 $$
which have solution
\begin{equation}  Z^\mu_\alpha    = \varphi^\nu_\alpha
{\partial{x^{\mu -}} \over \partial {x^{\nu -}_h}}\  ,\qquad
Z^{\mu ++}_\alpha  = \varphi^\nu_\alpha
{\partial{x^{\mu +}} \over \partial {x^{\nu -}_h}}\  .\end{equation}
By construction  (\ref{45}), as functions of the central frame
coordinates $\{ x^{\mu a}= x^{\mu+}u^{-a}-x^{\mu-}u^{+a} , u^\pm_a \}$,
these are bilinear in the harmonics, affording immediate extraction
of the vierbeins $E^{\mu b}_{\alpha a} =E^{\mu b}_{\alpha a}(x^{\nu c})$
in the customary space.

H. The connection $\omega_{\alpha}^+$ is given in terms of the bridge
by  the formula (\ref{16}), which therefore yields $\omega_{\alpha a} =
\omega_{\alpha a}(x^{\mu b})$, since $\omega_{\alpha}^+$, as a function
of central frame coordinates $\{ x^{\mu a} , u^\pm_a \}$, is by
construction (see (\ref{4})) linear in the harmonics $u^{+ a}$.

Therefore, extract
the self-dual  spin connection from the central frame formula
\begin{eqnarray} (\omega_{\alpha}^+ )^\gamma_\beta  &=&
(\omega_{\alpha a})^\gamma_\beta u^{+ a}  =
- D^+_\alpha \varphi^{\breve\beta}_\beta (\varphi^{-1})^\gamma_{\breve\beta}
\cr &=& - \varphi^{{\breve\gamma}}_\alpha D^+_{{\breve\gamma}}
        \varphi^{\breve\beta}_\beta (\varphi^{-1})^\gamma_{\breve\beta}  =
 - \varphi^{{\breve\gamma}}_\alpha
{\partial{ \varphi^{\breve\beta}_\beta} \over
\partial {x^{{\breve\gamma} -}_h}} (\varphi^{-1})^\gamma_{\breve\beta}
\cr &=& - ( Z^{++\mu}_\alpha {\partial{\varphi^{\breve\beta}_\beta } \over
\partial {x^{\mu +}}} + Z^\mu_\alpha
{\partial{\varphi^{\breve\beta}_\beta } \over \partial {x^{\mu -}}})
(\varphi^{-1})^\gamma_{\breve\beta}\      .\label{49}\end{eqnarray}

The thus constructed connection $(\omega_{\alpha a})^\gamma_\beta $ is
also a solution of an SU(2) self-dual  Yang-Mills  theory in a curved
space with metric (\ref{48}). This is obvious in our formulation since
this $\omega^+_\alpha$, by construction, satisfies (5a,b), which for this
connection are precisely the {\it Yang-Mills} CR conditions in a self-dual
background. This formulation therefore makes manifest the observation of
\cite{conn} that the spin connection $(\omega_{\alpha a})^\gamma_\beta$
corresponding to a self-dual  gravitational solution is such a self-dual
Yang-Mills vector potential.

\section{An example}\label{ex}

We now explicitly illustrate the procedure outlined above for the simplest
example of ${\cal L}^{+4}$: a monomial of fourth degree in the holomorphic
coordinates:
\begin{equation} {\cal L}^{+4} = g x^{1+}_h x^{1+}_h x^{2+}_h x^{2+}_h
   ,\label{50}\end{equation}
where $g$ is a dimensionful parameter.
This is invariant under the symmetry transformation
\begin{equation} x_h^{1+'}=e^{\gamma}x_h^{1+}, \quad
x_h^{2+'}=e^{-\gamma}x_h^{2+}  ,\label{51}\end{equation}
which plays an important role in the explicit solubility of this example.
The simplest quantity invariant under this symmetry is
$\rho^{++} \equiv x_h^{1+} x_h^{2+}$,
in terms of which this choice of ${\cal L}^{+4}$ allows expression.

Step A. From this prepotential, we get
the harmonic vielbeins $H^{++\mu +}$ using (\ref{41})
\begin{equation} H^{++\mu+}= \epsilon^{\mu\nu}
{\partial{{\cal L}^{+4}} \over \partial {x^{+\nu}_h}}
=2g \rho^{++} (\sigma_3 )_{\nu}^\mu x^{\nu+}_h
,\label{52}\end{equation} where $\sigma_3$ is the Pauli matrix. Using
these expressions, we find from (\ref{43}) that
\begin{equation} H^{++\mu-}
= x_h^{-{\breve\alpha}} \partial^-_{h {\breve\alpha}} H^{++{\breve\beta} +}
= 2g (\sigma_3 )_{\nu}^\mu  (x^{\nu-}_h\rho^{++}
+ x^{\nu+}_h (x^{1-}_h x^{2+}_h + x^{1+}_h x^{2-}_h) )
\label{53}\end{equation}
It is worth emphasizing once again that $H^{++\mu -}$ are non-analytic,
they contain the antiholomorphic coordinates $x_h^{\mu -}$ explicitly.

Step B. From the definition (\ref{13}) and having the explicit form for
$H^{++\mu+}$ (\ref{52}) we
can write down the equations for the holomorphic coordinates $x^{\mu +}_h$:
\begin{equation} \partial^{++}x^{\mu+}_h  =
2g \rho^{++} (\sigma_3 )_{\nu}^\mu x^{\nu+}_h\  .\label{54}\end{equation}
It follows from these equations that the invariant
$\rho^{++}$
is actually the conserved current corresponding to the symmetry (\ref{51}),
in the sense that
\begin{equation} \partial^{++} \rho^{++}=0 .\end{equation}
Due to this conservation law, eqs.(\ref{54}) are effectively {\it linear}
equations having solutions
\begin{equation} x^{1+}_h = e^{ 2g\rho} x^{1+},\quad
x^{2+}_h = e^{-2g\rho} x^{2+},\label{55} \end{equation}
where $\rho$ and $x^{\mu +}$ are solutions of the equations
\begin{eqnarray} &\partial^{++}\rho &= \rho^{++}\  ,\label{56}\\[0pt]
& \partial^{++}x^{\mu +}&=0\  \label{centr}.\end{eqnarray}
The latter equation allows the natural identification of the central
frame coordinates from the harmonic expansion $x^{\mu+}=x^{\mu a} u^+_a$;
and using (\ref{55}) $\rho^{++}$ may be seen to have the same form in
these central coordinates as in holomorphic ones, i.e.
$\rho^{++} = x^{1+}_h x^{2+}_h = x^{1+} x^{2+} $. Now (\ref{56}) may be
seen to have the following solution in terms of central coordinates:
$$ \rho= {1\over 2} (x^{1+} x^{2-} + x^{1-} x^{2+}) ,$$
up to the addition of a solution of the homogeneous part of (\ref{56}),
which can clearly be absorbed by redefinition of $x^{+\mu}$, which also
satisfies a homogeneous equation, (\ref{centr}).
The expressions (\ref{55}) are therefore indeed the required ones for
the holomorphic coordinates $x_h^{\mu+}$ in terms of the central ones.

Step C. The equations for the negatively charged coordinates are given
by (\ref{14}), which we solve using the positively charged coordinates
already found above. Inserting the explicit expressions (\ref{53}) we have
$$ \partial^{++} x^{1-}_h =
(1 +2g x^{1+}_h x^{2-}_h + 4g x^{1-}_h x^{2+}_h ) x^{1+}_h ,$$
\begin{equation} \partial^{++} x^{2-}_h =
(1 - 2g x^{1-}_h x^{2+}_h - 4g x^{1+} x^{2-}_h) x^{2+}_h\  ,\label{58}
\end{equation} equations linear in $x^{\mu -}_h$ as they stand.
Moreover, together with (\ref{55}), they imply that
$$ \partial^{++} (x^{1-}_h x^{2+}_h + x^{1+}_h x^{2-}_h)
= 2 x^{1+}_h x^{2+}_h ,$$ so comparing with (\ref{56}) and using the
linearity of $x^{\mu \pm}$ in the harmonics, we may make the identification
\begin{equation}  \rho= {1\over 2} (x^{1+} x^{2-} + x^{1-} x^{2+})
= {1\over 2} (x^{1+}_h x^{2-}_h + x^{1-}_h x^{2+}_h )  .\label{59}
\end{equation} So $\rho$, like $\rho^{++}$, has the same form in
both coordinate systems. We may now present the
equations in the form of a system linear in the holomorphic coordinates,
$$\partial^{++} x^{1-}_h
= x^{1+}_h + 4g \rho x^{1+}_h + 2g \rho^{++} x^{1-}_h\  ,$$
$$\partial^{++} x^{2-}_h
= x^{2+}_h - 4g \rho x^{2+}_h - 2g \rho^{++} x^{2-}_h\,$$ having solution:
\begin{equation}  x^{1-}_h = x^{1-} (1 + 2g x^{1+}x^{2-}) e^{2g\rho}, \quad
x^{2-}_h = x^{2-}(1 - 2g x^{1-} x^{2+})e^{-2g\rho} .\label{60}\end{equation}
Now since $\rho$ is the same in both central and holomorphic
coordinates, the relationships (\ref{55}) and (\ref{60}) are readily
invertible, yielding the following expressions for the central
coordinates as functions of the holomorphic ones:
$$x^{1+} = e^{-2g\rho} x_h^{1+},\quad x^{2+} = e^{2g\rho} x_h^{2+}$$
$$x^{1-} = e^{-2g\rho}(x_h^{1-} - 2g \rho^{--} x_h^{1+})
,\quad x^{2-} = e^{2g\rho}(x_h^{2-} + 2g \rho^{--} x_h^{2+}),$$
where $$\rho^{--} \equiv x^{1-}x^{2-}
= { 2 x^{1-}_h x^{2-}_h \over (1-g r_h^2 + \sqrt{1-2g r_h^2+16g^2\rho^2})};$$
and $r_h^2 \equiv 2(x_h^{1-} x_h^{2+} - x_h^{1+}x_h^{2-}).$

Step D. Using (\ref{40}) we find the connection $\omega^{++}$
to have the form
$$\omega^{++} = \partial^-_{h {\breve\alpha}} H^{++{\breve\beta} +}
= e^{-2g\sigma_3 \rho} \hat\omega^{++} e^{2g\sigma_3 \rho}$$
where
$$ \hat\omega^{++} = 2g \pmatrix {
2 x^{1+} x^{2+} & - x^{2+} x^{2+} \cr
  x^{1+} x^{1+} & -2 x^{1+} x^{2+} \cr}
\equiv 2g ( \sigma_3 \rho^{++} + X^{++}) ,$$
where $X^{++}$ is a matrix defined in appendix \ref{B}.

Step E. To find the bridge it is useful to present it in the manner,
$$\varphi = \hat\varphi e^{2g\sigma_3 \rho},$$ where $\hat \varphi$
satisfies, in virtue of (\ref{44}),
the equation  $$\partial^{++} \hat\varphi = 2g \hat\varphi X^{++} .$$
Now inserting the ansatz $\hat \varphi = e^{f(r^2)X^{+-}}$, where
$r^2 \equiv 2(x^{1-} x^{2+} - x^{1+}x^{2-})$ and $X^{+-}$ satisfies
$\partial^{++} X^{+-} =X^{++}$ and is given explicitly in appendix \ref{B},
into the relation
$X^{++}  = {1\over 2g} \hat\varphi^{-1} \partial^{++} \hat\varphi$,
the function $f(r^2)$ is determined to be
$f(r^2) = - {ln(1 - g r^2) \over g r^2}$, yielding
\begin{equation} \varphi = (1 - g r^2)^{-{1\over 2}}(1 + 2g X^{+-})
                          e^{2g\sigma_3 \rho} ,\label{62}\end{equation}
a result which may also be obtained (and easily checked) simply by linear
algebra  from  a careful consideration of the algebra (\ref{61}) in
appendix \ref{B}. The bridge $\varphi$ is defined by (\ref{44}) up to
multiplication by a factor whose $\partial^{++}$ derivative vanishes and
the unimodularity requirement,
$\det \varphi = 1$, yields the normalisation in (\ref{62}).

In virtue of the algebra of the matrices $X^{\pm\mp},X^{++}$ described in
appendix \ref{B}, the inverse bridge may immediately be written down:
$$\varphi^{-1} =
(1 - g r^2)^{-{1\over 2}}e^{-2g\sigma_3 \rho} (1 - 2g X^{-+})\   .$$

Step F. It follows from formulae (\ref{58}) and (\ref{60}) that
$${\partial{x^{\nu +}} \over \partial {x^{\mu -}_h}}
= - g e^{-2g\rho\sigma_3}  X^{++} $$ and that
$${\partial{x^{\nu -}} \over \partial {x^{\mu -}_h}} =
{ e^{-2g\rho\sigma_3} \over (1 - g r^2) }
         \left( 1 - g (3 - g r^2) X^{-+}\right)\  .$$

Step G. Substituting the latter expression and (\ref{62}) in (\ref{46})
we obtain:  $$Z^\mu_\alpha
=  \varphi^\nu_\alpha {\partial {x^{\mu -}} \over \partial {x^{\nu -}_h}}
=(1 - g r^2 )^{-{1\over 2}} (1 - g X^{-+})\   .$$
As expected, the harmonic expansion of $Z^\mu_\alpha$ contains only bilinear
pieces and the self-dual  vierbein consequently has the form
$$ E^{\mu b}_{\alpha a} =
(1 - g r^2 )^{-{1\over 2}} \pmatrix{
\delta^b_a - g x_a^2 x^{1b}  &  g x_a^2 x^{2b}  \cr
 - g x_a^1 x^{1b}            & \delta^b_a +  g x_a^1 x^{2b} \cr}$$
having inverse
$$ E_{\mu a}^{\alpha b} =
(1 - g r^2 )^{-{1\over 2}} \pmatrix{
\delta^b_a(1 - g r^2 ) + g x_a^2 x^{1b}  & - g x_a^2 x^{2b}  \cr
  g x_a^1 x^{1b}        & \delta^b_a(1 - g r^2 ) -  g x_a^1 x^{2b} \cr}.$$
Inserting this in (\ref{48}), using the parametrisation
$x^{\mu a} = \pmatrix{ y & -\bar z  \cr   z & \bar y \cr}$
and denoting $r^2  = 2(y\bar y + z\bar z)$, we find
\begin{equation} ds^2 = 2 (1 - gr^2 ) (dy d\bar y + dz d\bar z)
 + g r^4 {(2-gr^2)\over(1-gr^2)} \sigma_z^2\  ,\label{63}\end{equation}
where $\sigma_z$ is one of the three differential 1-forms \cite{eh}
related to the Maurer-Cartan forms on SU(2),
$$\begin{array}{rcl}
\sigma_x &=&{1\over r^2} (\bar z dy - y d\bar z + z d\bar y - \bar y dz),\cr
\sigma_y &=&{1\over r^2}(zdy - ydz +\bar y d\bar z - \bar z d\bar y),\cr
\sigma_z &=& {1\over r^2}(\bar y dy - y d\bar y + \bar z dz - z d\bar z).
\end{array} $$
Setting $g=0$ in (\ref{63}) (corresponding to ${\cal L}^{+4}=0$) we clearly
obtain the flat metric
\begin{equation} ds^2_{flat} = 2 (dy d\bar y + dz d\bar z) =
dr^2 + r^2 (\sigma_x^2 + \sigma_y^2 + \sigma_z^2) .\label{flat}\end{equation}
Using the second equality in (\ref{flat}) we obtain precisely the form of
the metric obtained using the harmonic space  formulation of N=2 sigma
models \cite{cmp}, viz. \begin{equation} ds^2 =
(1 - gr^2 ) dr^2  + r^2 (1 - gr^2 )(\sigma_x^2 + \sigma_y^2) +
{r^2\over(1-gr^2)} \sigma_z^2\  .\label{TN}\end{equation}
This is a form of the self-dual  Euclidean Taub-NUT metric. Denoting the
parameter $g = {1\over 4m^2}$, the variable change
$r^2 \rightarrow 2m(\rho-m)$
yields the form of the Taub-NUT metric in  e.g.\cite{eh}:
$$ds^2 = {\rho+m \over 4(\rho-m)} d\rho^2 +
(\rho^2-m^2)(\sigma_x^2 + \sigma_y^2)
+ 4m^2 {\rho-m \over \rho+m} \sigma_z^2\  .$$
Unlike (\ref{TN}), the latter form of the metric is only defined in the
domain $\rho > m $; and therefore does not have a well-defined flat limit
($m\rightarrow \infty$).

Step H. Substituting  derivatives of the central frame coordinates and
bridge $\varphi$ from steps E and F into (\ref{49}) we obtain
$$ (\omega^+_1 )^\gamma_\beta = - 2g (1-gr^2 )^{-{3\over2}}
                \pmatrix{ x^{2+}(1- {gr^2\over 2})& 0\cr
            x^{1+} & -x^{2+}(1- {gr^2\over 2})\cr}  $$
$$ (\omega^+_2 )^\gamma_\beta = - 2g (1-gr^2 )^{-{3\over2}}
                \pmatrix{ x^{1+}(1- {gr^2\over 2})& -x^{2+}\cr
             0 & -x^{1+}(1- {gr^2\over 2})\cr}\  ,$$
expressions manifestly linear in the harmonics, allowing us to extract the
connection components (\ref{3}) immediately:
$$(\omega_{\alpha a} )^\gamma_\beta = 2g (1-gr^2 )^{-{3\over2}}
        \pmatrix{ (1- {gr^2\over 2}) (x^{2}_a,x^{1}_a)& (0 ,  - x^{2}_a )\cr
	(x^{1}_a,   0   ) & -(1- {gr^2\over 2}) (x^{2}_a,x^{1}_a) \cr} .$$

\section{Conclusions and further remarks}\label{concl}

We have shown that {\it all} self-dual  gravitational fields allow
local description in terms of an unconstrained analytic prepotential
${\cal L}^{+4}$ in harmonic space .
The explicit performance of our construction relies only on the solution
of first order differential equations on $S^2$. Our method therefore
promises to be a fruitful one for the explicit construction of
self-dual  metrics.

Whether global characteristics (e.g. topological invariants) and singularity
properties of the manifold can be {\it determined} by a prescient choice
of ${\cal L}^{+4}$ remains an open question. The curvature tensor, however,
may indeed be evaluated from the analytic frame connection
$\omega^-_{\breve\beta}$ in virtue of  \begin{equation}
R_{{\breve\alpha}{\breve\beta}} =
\partial^+_{h({\breve\alpha}} \omega^-_{{\breve\beta})}\   ,\label{A6}
\end{equation} or equivalently
$$ C_{{\breve\alpha}
{\breve\beta}{\breve\gamma}}^{\;\;\;\;\;\;\; {\breve\delta}}
= \partial^+_{h({\breve\alpha}}
     \omega^{-\;\;{\breve\delta}}_{{\breve\beta}){\breve\gamma}}
= \partial^+_{h{\breve\alpha}} \partial^+_{h{\breve\beta}}
                      e^{--{\breve\delta}}_{\breve\gamma}\   ,$$
where the connection $\omega^-_{\breve\alpha}$ is determined in terms of
the vielbein $e^{--\mu}_{\breve\beta}$, which in turn is determined in
terms of ${\cal L}^{+4}$ (see appendix \ref{A}). The expression (\ref{A6}),
which is by construction u-independent, immediately yields the manifestly
total-derivative form of the Pontryagin density:
$$  R^{{\breve\alpha}{\breve\beta}}R_{{\breve\alpha}{\breve\beta}}
= \partial_h^{+{\breve\alpha}}  \partial_h^{+{\breve\beta}}
e^{--{\breve\delta}}_{\breve\gamma} \partial^+_{h{\breve\alpha}}
\partial^+_{h{\breve\beta}} e^{--{\breve\gamma}}_{\breve\delta}
= \partial_h^{+{\breve\alpha}} ( \partial_h^{+{\breve\beta}}
e^{--{\breve\delta}}_{\breve\gamma} \partial^+_{h{\breve\alpha}}
\partial^+_{h{\breve\beta}} e^{--{\breve\gamma}}_{\breve\delta}).$$

In our framework the fields $\omega^-_{\breve\alpha}$ and
$e^{--\mu}_{\breve\beta}$ are redundant, carrying no dynamical information;
all their equations of motion being identically satisfied in virtue of what
we have called the `dynamical' subset of CR equations.
Alternatively, one could choose to ignore the relations in (5)
involving ${\cal D}^{++} $ and attempt instead to solve the equations
for $e^{--\mu}_{\breve\beta}$, which describes the same degrees of freedom.
These equations imply the form (see (\ref{A5}) in Appendix \ref{A})
$e_{\breve\beta}^{--{\breve\gamma}}
= \partial^+_{h{\breve\beta}} \partial_h^{+{\breve\gamma}} e^{-4}$,
where $e^{-4}$ is a nonanalytic prepotential, which, in this alternative
framework, carries all the dynamical information. Indeed the equation
$$ [{\cal D}^-_\alpha, {\cal D}^-_\beta] = 0 ,$$
which completes the algebra of covariant differential operators in (5),
and which needs to be included if one wishes to exclude (5b,d)
(this equation is an identity in our framework), may easily be seen
to contain the dynamical equation for $e^{-4}$, \begin{equation}
\partial^{{\breve\alpha} +} \partial^-_{\breve\alpha} e^{-4} +
 {1\over 2} \partial^{{\breve\alpha} +}\partial^{{\breve\beta} +} e^{-4}
          \partial^+_{\breve\alpha} \partial^+_{\breve\beta} e^{-4} = 0\
         ,\label{A7} \end{equation}
as a consequence of the vanishing of the torsion coefficient of
$\partial^+_{h {\breve\alpha}}$. (These torsion constraints actually imply
that the left-hand side of (\ref{A7}) is equal to some arbitrary analytic
function, which however may be set to zero using a pregauge symmetry of
$e^{-4}$, viz. $ \delta e^{-4} = x^{\mu -} G_\mu^{-3}(x^+)$, for arbitrary
analytic $G_\mu^{-3}$). In the alternative approaches to the self-dual
Einstein equations which do not introduce the auxiliary ${\cal D}^{++} $
(e.g. \cite{pleb,grant}), the dynamics is indeed described by (\ref{A7})
or transformations thereof. The field $e^{-4}$ is precisely Plebanski's
second `heavenly function' \cite{pleb}    \footnote{The conjugation of
harmonics which yields real $x^\pm$ of course does not hold if the harmonics
are given fixed numerical values and (\ref{A7}) is then an equation in
${\kern2pt {\hbox{\sqi I}}\kern-4.2pt\rm C}^4.$}. Indeed the
contravariant form of the basis (\ref{h-deriv}) with Cartan 1-forms
$$\begin{array}{rcl}  \Omega^{\mu +} &=&  dx_h^{\mu +} \cr
\Omega^{\mu -} &=& dx_h^{\mu -} + e^{--\mu}_\nu dx_h^{\nu +} \end{array} $$
yields the (u-independent) invariant
$ dx_{h\nu}^+ (dx_h^{\nu -} + e^{--\nu}_\mu dx_h^{\mu +}) ,$
which is precisely the form of the metric given in \cite{bfp}. We emphasise
however that the advantage of introducing  ${\cal D}^{++} $ and working
with harmonics as independent variables, is that eq. (\ref{A7}) is then
automatically satisfied. If explicit solutions of (\ref{A7}) are for some
reason required, they may also be constructed in our framework from a
first order equation ((\ref{A2}) in Appendix A).

The alternative Monge-Amp\`ere form of Plebanski's equation \cite{pleb}
corresponds to a different gauge to (\ref{h-flat}), i.e. a different
choice of analytic frame coordinates; and our method may be used to
construct solutions to that form of Plebanski's equation as well.
We hope to return to a discussion of both of Plebanski's equations
in the harmonic space  setting in a future publication.

The self-duality  conditions are well known to be differential equations
whose solutions are automatically hyper-K\"ahler metrics. Our construction
of solutions generalises to $4m$-dimensions, where hyper-K\"ahler metrics
\cite{GIOS} may similarly be thought of as solutions to the generalised
self-duality  conditions \cite{ward3}
\begin{equation} [{\cal D}_{B b}, {\cal D}_{A a}] =
                  \epsilon_{a b} R_{AB},\label{hk}\end{equation}
where $A$ is an $Sp(m)$ index and $a$, as presently, an Sp(1) index.
These equations manifestly break the $4m$-dimensional rotation group
to $Sp(m)\times Sp(1)$. Delocalising the $Sp(1)$, yields connections and
curvatures manifestly taking values in the $Sp(m)$ subalgebra. In other
words the holonomy group is $Sp(m)$, so eqs. (\ref{hk}) indeed describe
higher dimensional hyper-K\"ahler spaces. Our entire construction
generalises to these higher dimensional cases on replacing the $Sp(1)$
indices $\alpha,\beta$ by $Sp(m)$ indices $A,B$. The further generalisation
with indices $A,B$ in (\ref{hk}) considered as `superindices' of the
superalgebra $OSp(N|m)$, yields constraints in chiral superspace which
may be thought of as equations for N-extended supersymmetric
$(4m|2N)$-dimensional hyper-K\"ahler spaces. In this supersymmetric case,
our construction requires suitable modification in order to accommodate
intricacies of superalgebras. The case $m=1$ corresponds to N-extended
supersymmetric self-dual  supergravity. In fact, the present work was
motivated by our initial attempts to supersymmetrise the harmonic-twistor
construction for self-dual  gravity and the construction
of this paper is indeed amenable to supersymmetrisation, yielding a general
solution of all extended self-dual  Poincar\'e supergravity theories.
This is under preparation for publication.

We are grateful to A. Galperin and E. Ivanov for useful discussions.
One of us (V.O.)
is happy to thank the Physikalisches Institut, Universit\"at Bonn
and the Max-Planck-Institut f\"ur Mathematik, Bonn for support and
hospitality during the performance of the early stages of this work.
This paper was completed during a visit to Vienna. We thank the
Erwin Schr\"odinger Institute for support and hospitality.
\begin{appendix}
\renewcommand\theequation{\thesection{.\arabic{equation}}}
\appendix
\section{Appendix. The redundancy of the degrees of freedom in
${\cal D}^-_\alpha$ } \label{A}
\setcounter{equation}{0}
In this appendix we prove the claim in section \ref{solution} that the
fields $e^{--\mu}_{\breve\beta}$ and $\omega^-_{\breve\alpha}$, are
redundant degrees of freedom and are indeed entirely determined in terms
of the `dynamical' fields $H^{++\mu+}$ and $\omega^{++}$, which in turn
are determined by the analytic prepotential ${\cal L}^{+4}$.

Equation (\ref{34}) reduces to
\begin{equation}  {\cal D}^{++} e^{--\mu}_{\breve\alpha} =
                - \partial^-_{h {\breve\alpha}} H^{++ \mu -}
+e^{--\nu}_{\breve\alpha} \partial^+_{h \nu} H^{++ \mu -} .\label{A1}
\end{equation}
Now recalling the first expression for $\omega^{++}$ in (\ref{42}), we
note that this equation is actually just the equation
$$ {\cal D}^{++} e^{--{\breve\beta}}_{\breve\alpha} =
      - \partial^-_{h {\breve\alpha}} H^{++ {\breve\beta} -}$$
with the connection acting in the gauge (\ref{h-flat}) on {\it both}
spinor indices of $e^{--{\breve\beta}}_{\breve\alpha}$.
Now in the central basis this equation reads simply
\begin{equation}  \partial^{++} e^{--\beta}_\alpha
= - \varphi^{\breve\alpha}_\alpha \partial^-_{h {\breve\alpha}}
H^{++ {\breve\beta} -}(\varphi^{-1})_{\breve\beta}^\beta,\label{A2}
\end{equation}
a first order linear equation which uniquely determines $e^{--\beta}_\alpha$
in terms of $\varphi$ and $H^{++ \beta -}$, which in turn are determined in
terms of the arbitrary analytic prepotential ${\cal L}^{+4}$ in virtue
of (\ref{40}, \ref{43}).

For the Taub-NUT example of section \ref{ex},
the explicit integration of (\ref{A2}) yields
\begin{equation} e^{--\beta}_\alpha = {4g\over (1-gr^2)}
\left( X^{--}\{ (1- {gr^2\over 2} )^2 - {1\over 2} - 4 g^2 \rho^2 ) \}
+\sigma_3({1\over 2} \rho^{--}I - 2g\rho X^{--}) + 2g\rho\rho^{--}I
\right)^\beta_\alpha  \
  ,\end{equation}  where $X^{--}$ is the matrix in (\ref{B2}).

{}From (\ref{35}) we obtain an expression for the connection in
${\cal D}^-_{\breve\beta}$:
\begin{equation} \omega^{- {\breve\gamma}}_{{\breve\beta}{\breve\alpha}} =
\partial^+_{h {\breve\alpha}} e_{\breve\beta}^{--{\breve\gamma}} ,\label{A4}
\end{equation}
for which (\ref{36}) implies the constraint
$\omega^{- {\breve\gamma}}_{[{\breve\beta}{\breve\alpha}]} = 0$.
The latter together with the constraint of tracelessness of this SU(2)
connection have local solution in terms of an unconstrained non-analytic
prepotential $e^{-4} = e^{-4}(x_h^\pm,u^\pm)$, in terms of which
\begin{equation} \begin{array}{lcr} e_{\breve\beta}^{--{\breve\gamma}} &=&
\partial^+_{h{\breve\beta}} \partial_h^{+{\breve\gamma}} e^{-4},\cr
 \omega^{- {\breve\gamma}}_{{\breve\beta}{\breve\alpha}} &=&
\partial^+_{h{\breve\alpha}} \partial^+_{h{\breve\beta}}
 \partial_h^{+{\breve\gamma}} e^{-4}\   .\label{A5}\end{array}\end{equation}

The only remaining equation for the connection fields is the equation
(\ref{37}), which is an identity in virtue of the other equations. Namely,
acting on (\ref{A1}) by $\partial^+_{\breve\beta}$ we obtain precisely
eq. (\ref{37}) with the connection components written in the forms
(\ref{42}) and (\ref{A4}).

All equations in the CR system therefore allow solution in terms of the
unconstrained analytic prepotential ${\cal L}^{+4}$, proving our claim.

\section{Appendix. The quadratic matrices $X^{\pm\pm},X^{\pm\mp}$}\label{B}
\setcounter{equation}{0}
The explicit solubility of the example of section \ref{ex} relies on
remarkable properties of  matrices quadratic in the central coordinates
$x^{\mu \pm}\equiv x^{\mu a} u^\pm_a$.  Consider
$$ X^{++} =   \pmatrix{
 x^{1+} x^{2+} &  -  x^{2+}x^{2+} \cr
 x^{1+} x^{1+} &  -  x^{2+} x^{1+} \cr }.$$
This is actually the harmonic derivative of two possible matrices,
$X^{+-}, X^{-+}$ ,
$$X^{++}=\partial^{++} X^{+-}=\partial^{++} X^{-+}\  ;\quad
X^{+-}=  \pmatrix{
x^{1+} x^{2-} & - x^{2+} x^{2-} \cr
x^{1+} x^{1-} & - x^{2+} x^{1-} \cr},
X^{-+} =  \pmatrix{
x^{1-} x^{2+} & - x^{2-}x^{2+} \cr
x^{1-} x^{1+} & - x^{2-} x^{1+} \cr} ,$$
together with which it obeys the algebra
\begin{equation}\begin{array}{rl}
& X^{++} X^{++} = X^{+-}X^{++} = X^{++} X^{-+} = 0 \cr
& X^{+-} X^{-+} = X^{-+} X^{+-} = 0 \cr
& X^{++} X^{+-} = - {r^2\over 2} X^{++} = - X^{-+} X^{++} \cr
& X^{+-} X^{+-} = - {r^2\over 2} X^{+-}\  ,\qquad
  X^{-+} X^{-+} =   {r^2\over 2} X^{-+} \cr
& X^{+-}- X^{-+} = - {r^2\over 2} I  .\end{array}\label{61}\end{equation}
where  $r^2 \equiv 2(x^{1-} x^{2+} - x^{1+}x^{2-}).$
The further matrix
\begin{equation} X^{--} =   \pmatrix{
 x^{1-} x^{2-} &  -  x^{2-}x^{2-} \cr
 x^{1-} x^{1-} &  -  x^{2-} x^{1-} \cr }\label{B2}\end{equation}
satisfies the equation $\partial^{++} X^{--} = X^{+-} + X^{-+}$.
\end{appendix}

\end{document}